\documentclass[%
preprint,
 amsmath,amssymb,
 aps,
 pre,
]{revtex4-1}

\usepackage{mathrsfs} %ADDED BY ME

\usepackage{graphicx}% Include figure files
\usepackage{dcolumn}% Align table columns on decimal point
\usepackage{bm}% bold math
\usepackage[table,xcdraw]{xcolor}
\usepackage{adjustbox}

\newcommand{\pd}[3][]{\frac{\partial^{#1} #2}{\partial #3^{#1}}}
\usepackage{amsmath}
\DeclareMathOperator{\csch}{csch}
\begin{document}

%\preprint{APS/123-QED}

\title{Wall effects of eccentric spheres machine learning for convenient computation}
%\title{Improving Particle Tracking Microrheometry Using Non-linear Driving Forces}% Force line breaks with \\

\author{Lachlan J. Gibson}
\author{Shu Zhang}%
\author{Alexander B. Stilgoe}
\author{Timo A. Nieminen}
\author{Halina Rubinsztein-Dunlop}
\affiliation{%
The University of Queensland, School of Mathematics and Physics, Brisbane QLD 4072, Australia
}%

\date{\today}% It is always \today, today,

\begin{abstract}
In confined systems, such as the inside of a biological cell, the outer boundary or wall can affect the dynamics of internal particles. In many cases of interest both the internal particle and outer wall are approximately spherical. Therefore, quantifying the wall effects from an outer spherical boundary on the motion of an internal eccentric sphere is very useful. However, when the two spheres are not concentric, the problem becomes non-trivial. In this paper we improve existing analytical methods to evaluate these wall effects and then train a feed-forward artificial neural network within a broader model. The final model generally performed with $\sim0.001\%$ error within the training domain and $\sim0.05\%$ when the outer spherical wall was extrapolated to an infinite plane. Through this model, the wall effects of an outer spherical boundary on the arbitrary motion of an internal sphere for all experimentally achievable configurations can now be conveniently and efficiently determined.
%\begin{description}
%use the optional argument of the \verb+\item+ command to give the category of %each item. 
%\end{description}
\end{abstract}

\pacs{Valid PACS appear here}% PACS, the Physics and Astronomy
\maketitle

%\tableofcontents

\section{Introduction}
Quantifying effects of boundaries on the dynamics and behaviour of microscopic entities in biological fluids is a problem intersecting several fields of research including microrheology \cite{Shu_2018}, optical tweezers \cite{Di} and microbiology \cite{Nosrati_2015,papavassiliou_alexander_2017}. In most scenarios evaluating these so-called wall effects is non-trivial. In cases where the wall effects have been solved analytically, the given expressions are often difficult or inconvenient to evaluate because of their large size or ill-behaviour. Therefore, this paper aims to make computing the wall effects of eccentric spheres simple and more efficient by improving various analytical results and training a neural network model to be able to efficiently replicate the analytical results.

Wall effects of eccentric spheres, where an outer spherical boundary affects the dynamics of an internal centre-offset sphere through hydrodynamic interactions, are important in a variety of applications. For example, the intracellular environments of living cells plays an important role in cellular and sub-cellular processes such as replication and intracellular trafficking \cite{Wirtz2009}. Some microrheological techniques rely on the dynamics of spherical probe particles \cite{Bennett,Zhang_2017}, which could be used to explore properties of the cytoplasm to help understand cellular mechanisms \cite{Gibson_2017}.

In general, measuring the dynamics of probe particles is a typical approach to determine mechanical properties of complex fluids \cite{Guo}. To make such measurements, one must not only detect the probe particles but also track their motion in local space. However, in some experiments which work in confined environments, such as inside the cell, the influences of the boundaries the motion of the probe become non-negligible.

This novel approach for accurate cellular rheology requires calibration factors of the hydrodynamic interaction between the probe and near boundary walls. The wall effects of an infinite plane on the translation and rotation of a sphere are well known \cite{Leach, Chaoui, Jeffery1915, DeanONeill_1963}. In cases of more complex systems, the hindered translational diffusion has been studied extensively for spherical particles moving between two plane walls \cite{Lin}. Furthermore, cylindrical geometries and linear channels were also studied in a few cases, such as measurements of the drag coefficient of a sphere settling along the axis \cite{Eral, Dettmer}. Zhang \textit{et al}.\cite{Shu} have recently measured the wall effects of an artificial liposome, which is approximately spherical, on the rotation of an internal spherical particle. To our knowledge, no study has evaluated the collective translational and rotational wall effects so that the drag forces acting on a sphere in arbitrary motion could be easily computed.

\section{Theory}\label{sec_theory}
To quantify the wall effects of an outer sphere on an internal sphere, the equations of motion of the fluid are evaluated. From the fluid velocity and pressure the torque and force acting on the internal sphere can be extracted. Comparing these values with the drag forces acting on a sphere in an open fluid reveals the effects of the outer wall on the rotation and translation of the internal particle.

Analytical methods to evaluate these wall effects have been established quite some time ago by Jeffery \cite{Jeffery_1912,Jeffery1915,Stimson_1926}, Stimson \cite{Stimson_1926}, Majumdar \cite{Majumdar_1969,ONeill_1970P1,ONeill_1970P2} and O'Neill \cite{ONeill_1970P1,ONeill_1970P2}. Summaries of their methods as well as novel improvements are presented in Appendices \ref{appendix:AxiRotWE}, \ref{appendix:AxiTraWE}, \ref{appendix:AsyRotWE} and \ref{appendix:AsyTraWE}. In section \ref{sec_MachineLearning} these analytically based methods will be used to generate training data for a neural network model to learn to replicate the analytical results.

\subsection{Problem Construction}
\subsubsection{Geometry and Bispherical Coordinates}
\begin{figure}
\centering
\includegraphics*[trim={1cm 0.5cm 1cm 0.6cm},clip,width=0.8\linewidth]{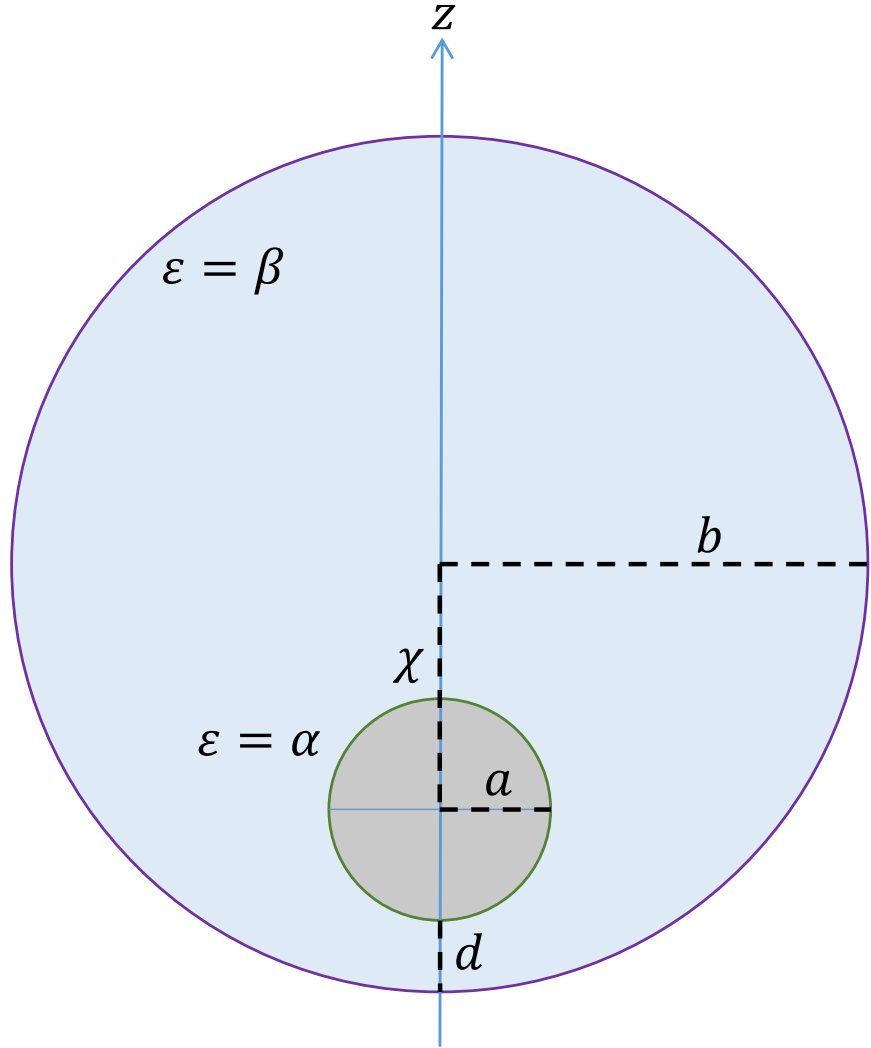}
\caption{The inner sphere, with radius $a$ is offset from the outer spherical boundary by $\chi$ along the $z$-axis. The minimum clearance distance between boundaries can be related to the radii and vertical offset by $d=b-a-\chi$.}\label{fig_geometry}
\end{figure}

Before evaluating any equations, a suitable coordinate system needs to be chosen to frame the problem. Typically in the case of eccentric spheres, where the sphere centres are offset, bispherical coordinates are the natural choice as they form an orthogonal coordinate system with eccentric spherical coordinate surfaces that lie along the $z$-axis. The bispherical coordinates $(\varepsilon,\theta,\psi)$ to cylindrical coordinates $(r,\theta,z)$ transformation is given by
\begin{equation}\label{eq_bispherical_coordinates}
r=\frac{c \sin\psi}{\cosh \varepsilon-\cos\psi}, \qquad z=\frac{c \sinh\varepsilon}{\cosh \varepsilon-\cos\psi},
\end{equation}
where $c$ is a parameter yet to be determined by the positions and radii of the two eccentric spheres. $r$, $\theta$ and $z$ are the standard cylindrical radial, azimuthal and vertical coordinates respectively. From these transformation equations we find that
\begin{equation}
r^2+(z-c \coth \varepsilon)^2=(c \csch\varepsilon)^2,
\end{equation}
demonstrating how the coordinate $\varepsilon$ parametrises the radius and $z$ position of the spherical coordinate surfaces by $c \csch\varepsilon$ and $c \coth \varepsilon$, respectively. Without loss of generality, the inner and outer spherical boundaries are set to reside at $\varepsilon=\alpha$ and $\varepsilon=\beta$, respectively. Figure \ref{fig_geometry} illustrates this configuration. Positioning the spheres along the positive $z$-axis requires $0\leqslant\beta<\alpha$ and so the fluid fills the region $\beta<\varepsilon<\alpha$. When $\beta=0$ the outer sphere becomes an infinite plane at $z=0$. Fixing the boundary radii ($a$ and $b$) and their centre offset ($\chi$) determines $c$ by,
\begin{equation}
c=\frac{\sqrt{(a^2-b^2+\chi^2)^2-4a^2\chi^2}}{2\chi}.
\end{equation}
In the special case of an infinite plane ($b\rightarrow\infty$), this equation reduces to,
\begin{equation}
c=\sqrt{d(2a+d)}
\end{equation}
where $d=b-a-\chi$ is the minimum clearance distance between the two boundaries.

\subsubsection{Equations of Motion}
The fluid dynamics are modelled using classical continuum equations. If mass is conserved and the fluid is incompressible then the continuity equation requires that the divergence of the fluid velocity $\mathbf{v}$ is zero,
\begin{equation}\label{eq_ContinuityEquation}
\nabla\cdot\mathbf{v}=0.
\end{equation}
If momentum is conserved then the Cauchy momentum equation relates the total force acting on an infinitesimal volume element to the sum of external forces $\mathbf{f_{ext}}$ and the fluid's stress tensor $\mathbf{\sigma}$,
\begin{equation}\label{eq_CauchyMomentum}
\rho\frac{D\mathbf{v}}{Dt}=\mathbf{f_{ext}}+\nabla\cdot\mathbf{\sigma}
\end{equation}
where $\rho$ is the fluid density and $\frac{D\mathbf{v}}{Dt}$ is the material derivative.

As would be expected in the relevant microscopic systems mentioned in the introduction, external forces acting on the fluid are assumed to be negligible. Similarly, in the low Reynolds number limit the inertial terms are neglected. So equation (\ref{eq_CauchyMomentum}) is reduced to $\nabla\cdot\mathbf{\sigma}=\mathbf{0}$.

Modelling the fluid as an isotropic Newtonian incompressible viscous fluid results in a symmetric stress tensor that depends on the pressure $p$ and dynamic viscosity $\eta$
\begin{equation}
\mathbf{\sigma}=-p\mathbf{I}+\eta\left(\nabla\mathbf{v}+\nabla\mathbf{v}^T\right)
\end{equation}
where $\mathbf{I}$ is the identity tensor and the $T$ superscript denotes transposition. Setting the divergences of this stress tensor and the fluid velocity to zero results in the Stokes equations,
\begin{equation}
\eta \nabla^2 \mathbf{v}=\nabla p, \qquad \nabla\cdot\mathbf{v}=0,
\end{equation}
which, together with appropriate boundary conditions fully model the fluid dynamics. The Stokes equations in cylindrical coordinates and vector components \cite{Majumdar_1969} are given by
\begin{align}
\nabla^2u-\frac{2}{r^2}\pd{v}{\theta}-\frac{u}{r^2}&=\frac{1}{\eta}\pd{p}{r},\label{eq_equation_of_motion1}\\
\nabla^2v+\frac{2}{r^2}\pd{u}{\theta}-\frac{v}{r^2}&=\frac{1}{\eta r}\pd{p}{\theta},\label{eq_equation_of_motion2}\\
\nabla^2w&=\frac{1}{\eta}\pd{p}{z},\label{eq_equation_of_motion3}\\
\pd{u}{r}+\frac{u}{r}+\frac{1}{r}\pd{v}{\theta}+\pd{w}{z}&=0,\label{eq_equation_of_motion4}\\
\nabla^2p&=0,\label{eq_equation_of_motion5}
\end{align}
where $u$, $v$ and $w$ are the standard cylindrical vector components and
\begin{equation}
\nabla^2=\pd[2]{}{r}+\frac{1}{r}\pd{}{r}+\frac{1}{r^2}\pd[2]{}{\theta}+\pd[2]{}{z}.
\end{equation}

\subsubsection{Boundary Conditions}\label{sec_BoundaryConditions}
\begin{figure}
\centering
\includegraphics*[width=0.8\linewidth]{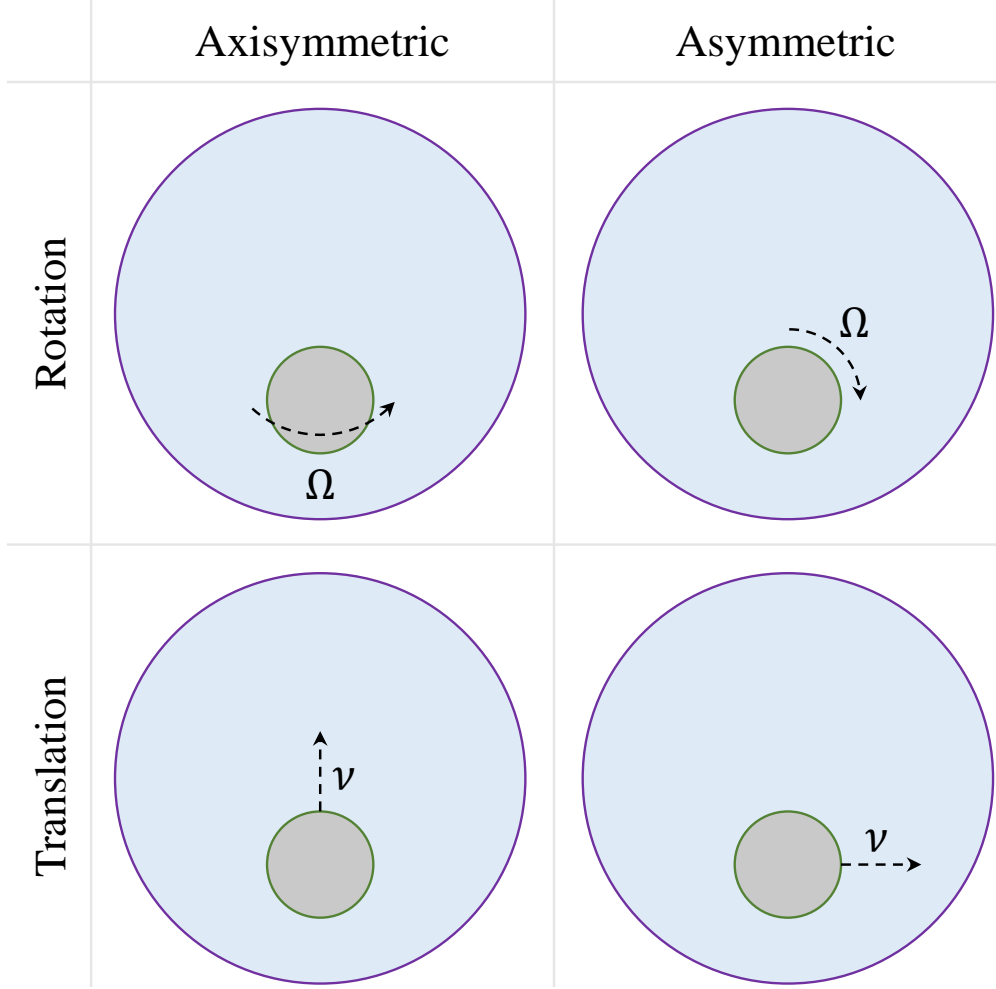}
\caption{The four distinct motions of the inner sphere. The top and bottom rows distinguish rotation and translation. The left and right columns distinguish axisymmetry and asymmetry.}\label{fig_four_motions}
\end{figure}

The linearity of the equations of motion means solutions can be expressed as linear combinations of other solutions. As a result, the problem of modelling arbitrary dynamics of the inner sphere (while the outer sphere is stationary) can be reduced to only four sub-problems. All kinds of motion from the inner sphere can be expressed as a linear combination of orthogonal rotations and translations. As illustrated in figure \ref{fig_four_motions}, the symmetry of the spheres allows for four cases: axisymmetric rotation, axisymmetric translation, asymmetric rotation and asymmetric translation. Therefore, the drag forces acting on the inner sphere can always be evaluated as a combination of these four cases.

It is assumed that the fluid follows stick boundary conditions whereby the fluid velocity at each boundary matches the corresponding boundary velocity. In all four cases the outer boundary is assumed to be stationary so all velocity components are zero when $\varepsilon=\beta$. The boundary conditions at the inner spherical boundary ($\varepsilon=\alpha$) in cylindrical vector components for axisymmetric and asymmetric rotation respectively are

\begin{equation}\label{eq_RotationalBCs}
\begin{bmatrix}
u\\
v\\
w
\end{bmatrix}
=\Omega
\begin{bmatrix}
0\\
r\\
0
\end{bmatrix},
\qquad
\begin{bmatrix}
u\\
v\\
w
\end{bmatrix}
=\Omega
\begin{bmatrix}
(z-z_0)\cos\theta\\
-(z-z_0)\sin\theta\\
-r\cos\theta
\end{bmatrix},
\end{equation}
where $\Omega$ is the angular velocity of the inner sphere and $z_0=c\coth\alpha$. The boundary conditions for axisymmetric and asymmetric translation at the inner sphere are
\begin{equation}\label{eq_TranslationalBCs}
\begin{bmatrix}
u\\
v\\
w
\end{bmatrix}
=\nu
\begin{bmatrix}
0\\
0\\
1
\end{bmatrix},
\qquad
\begin{bmatrix}
u\\
v\\
w
\end{bmatrix}
=\nu
\begin{bmatrix}
\cos\theta\\
-\sin\theta\\
0
\end{bmatrix},
\end{equation}
where $\nu$ is the linear velocity of the inner sphere.

%\subsubsection{Dimensionality Reduction}
%It is possible to reduce the number of spatial dimensions to two in all four cases. In both axisymmetric cases the boundary conditions are rotationally symmetric so the whole velocity and pressure fields do not depend on $\theta$. Therefore, the equations of motions simply to
%
%\begin{align}
%\nabla^2u-\frac{u}{r^2}&=\frac{1}{\eta}\pd{p}{r},\\
%\nabla^2v-\frac{v}{r^2}&=0,\\
%\nabla^2w&=\frac{1}{\eta}\pd{p}{z},\\
%\pd{u}{r}+\frac{u}{r}+\pd{w}{z}&=0,\\
%\nabla^2p&=0.
%\end{align}
%

\subsubsection{Drag Force and Torque}
Calculating the drag force and torque acting on the inner sphere involves evaluating the force and torque acting on surface elements of the sphere and then integrating over the whole surface. For a sphere centred on the $z$ axis at $z_0$, a position vector $\mathbf{r}$ from the centre to the surface can be expressed in cylindrical vector components as
\begin{equation}
\mathbf{r}=
\begin{bmatrix}
r\\0\\z-z_0
\end{bmatrix}.
\end{equation}
Therefore, the surface normal vector $\hat{n}$ (inward with respect to the particle but outward with respect to the fluid) is given by
\begin{equation}
\hat{n}=-\frac{\mathbf{r}}{a}.
\end{equation} 
The stress $\mathbf{P}$ acting on a surface element of the particle is the negative dot product of the unit normal vector and the stress tensor. In cylindrical components, the surface force density acting on the particle is
\begin{equation}\label{eq_SurfaceForceDensity}
\mathbf{P}=-\hat{n}\cdot \mathbf{\sigma}=\frac{1}{a}
\begin{bmatrix}
r\sigma_{rr}+(z-z_0)\sigma_{rz}\\
r\sigma_{r\theta}+(z-z_0)\sigma_{\theta z}\\
r\sigma_{rz}+(z-z_0)\sigma_{zz}
\end{bmatrix}
\end{equation}
where $\sigma_{ij}$ represent the stress tensor components in cylindrical coordinates \cite{Landau}
\begin{align}\label{eq_StressTensorCyl}
\begin{split}
\sigma_{rr}&=-p+2\eta\pd{u}{r},\\
\sigma_{\theta\theta}&=-p+2\eta\left(\frac{1}{r}\pd{v}{\theta}+\frac{v}{r}\right),\\
\sigma_{zz}&=-p+2\eta\pd{w}{z},\\
\sigma_{r\theta}&=\eta\left(\frac{1}{r}\pd{u}{\theta}+\pd{v}{r}-\frac{v}{r}\right),\\
\sigma_{rz}&=\eta\left(\pd{u}{z}+\pd{w}{r}\right),\\
\sigma_{\theta z}&=\eta\left(\pd{v}{z}+\frac{1}{r}\pd{w}{\theta}\right).
\end{split}
\end{align}

Therefore, the surface torque density $\mathbf{T}$ acting on the particle is
\begin{equation}\label{eq_SurfaceTorqueDensity}
\mathbf{T}=\mathbf{r}\times\mathbf{P}=
\begin{bmatrix}
-(z-z_0)P_\theta\\
(z-z_0)P_r-rP_z\\
rP_\theta
\end{bmatrix}
\end{equation}
where $P_r$, $P_\theta$ and $P_z$ are the cylindrical vector components of $\mathbf{P}$ as shown in equation (\ref{eq_SurfaceForceDensity}).

Evaluating the total force $\mathbf{F}$ and torque $\mathbf{G}$ acting on the particle involves integrating the force and torque densities over the whole spherical surface. In bispherical coordinates the surface integrals are
\begin{equation}\label{eq_ForceTotal}
\mathbf{F}=\int_{0}^{2\pi}\int_{0}^{\pi}\mathbf{P}\frac{c^2\sin\psi}{(\cosh\alpha-\cos\psi)^2}\, d\psi d\theta
\end{equation}

\begin{equation}\label{eq_TorqueTotal}
\mathbf{G}=\int_{0}^{2\pi}\int_{0}^{\pi}\mathbf{T}\frac{c^2\sin\psi}{(\cosh\alpha-\cos\psi)^2}\, d\psi d\theta.
\end{equation}

\subsubsection{Drag from Arbitrary Motion}
As will be explored in subsequent sections, the majority of the force and torque vector components for each kind of motion are zero, and the non-zero components are linear combinations of the particle velocity and rotation components. In general the total force and torque acting on the inner sphere can be described by \cite{HappelJohn1983LRnh}
\begin{align}
\mathbf{F}&=-\eta(\mathbf{K}\cdot\mathbf{V}+\mathbf{C}^T\cdot\mathbf{\Omega})\\
\mathbf{G}&=-\eta(\mathbf{C}\cdot\mathbf{V}+\mathbf{O}\cdot\mathbf{\Omega})
\end{align}
where $\mathbf{K}$ is the translational tensor, $\mathbf{O}$ is the rotational tensor, and $\mathbf{C}$ is the coupling tensor which describes the coupling between rotational and translational motions and forces. For eccentric spheres with centres lying on the $z$ axis, these tensors in Cartesian coordinates can be written in terms of dimensionless quantities $f_i$, $g_i$, $f_i^c$ and $g_i^c$:
\begin{align}
\mathbf{K}&=6\pi a
\begin{bmatrix}
f_x & 0 & 0\\
0 & f_y & 0\\
0 & 0 & f_z
\end{bmatrix} &
\mathbf{O}&=8\pi a^3
\begin{bmatrix}
g_x & 0 & 0\\
0 & g_y & 0\\
0 & 0 & g_z
\end{bmatrix}\\
\mathbf{C^T}&=6\pi a^2
\begin{bmatrix}
0 & f_x^c & 0\\
f_y^c & 0 & 0\\
0 & 0 & 0
\end{bmatrix} &
\mathbf{C}&=8\pi a^2
\begin{bmatrix}
0 & g_x^c & 0\\
g_y^c & 0 & 0\\
0 & 0 & 0
\end{bmatrix}.
\end{align}

When the centres of the eccentric spheres are both positioned on the $z$ axis, then the same asymmetric results can be applied to both the $x$ and $y$ dimensions giving the following relations
\begin{align}
f_y&=f_x, & g_x&=g_y, & f_y^c&=-f_x^c, & g_x^c&=-g_y^c.
\end{align}
Because of the Lorentz reciprocal theorem, the coupling tensors are related by a transpose \cite{HappelJohn1983LRnh} so a fifth condition is
\begin{equation}
g_y^c=\frac{3}{4}f_x^c.
\end{equation}

Therefore, for arbitrary translation and rotation in three dimensions the total force and torque vectors can be evaluated from just $f_x$, $f_z$, $g_y$, $g_z$ and $f_x^c$ which will be found from the axisymmetric translation, asymmetric rotation, axisymmetric rotation and asymmetric rotation respectively.

\subsection{Summary of Analytical Results}
Derivations of analytical expressions for $g_z$, $f_z$, $g_y$, $f_x$ and $f_x^c$, or related problems, have previously been established by Jeffery \cite{Jeffery_1912,Jeffery1915,Stimson_1926}, Stimson \cite{Stimson_1926}, Majumdar \cite{Majumdar_1969,ONeill_1970P1,ONeill_1970P2} and O'Neill \cite{ONeill_1970P1,ONeill_1970P2}. Improved versions of these derivations are included in Appendices \ref{appendix:AxiRotWE}, \ref{appendix:AxiTraWE}, \ref{appendix:AsyRotWE} and \ref{appendix:AsyTraWE}. This section outlines the final results and summarises our contributions.

\subsubsection{Axisymmetric Rotational Wall Effect}
The problem of finding the axisymmetric rotational wall effect has been previously solved analytically by Jeffery\cite{Jeffery1915} where he used a series solution to solve the equations of motion. From that solution he produced two separate series expressions for $g_z$, equations \eqref{eq_AxiRSeries1} and \eqref{eq_AxiRSeries2}. Appendix \ref{appendix:AxiRotWE} presents an outline of a very similar derivation of these solutions. We have managed to merge the two series into a single expression \eqref{eq_AxiRSeries3} that converges much more quickly than either component individually,
\begin{align}
\begin{split}
g_z&=\sum^{M}_{m=0}\left(\frac{\sinh\alpha}{\sinh(\alpha+m(\alpha-\beta))}\right)^3\\
&+4\sinh^3\alpha\sum^\infty_{n=1}\frac{n(n+1)e^{-(M+1)(2n+1)(\alpha-\beta)}}{e^{(2n+1)\alpha}-e^{(2n+1)\beta}}.
\end{split}
\end{align}

Jeffery\cite{Jeffery1915} failed to produce any expression for the wall effects in the low clearance limit, $d\rightarrow0$. We have achieved this by taking this limit of the summand in equation \eqref{eq_AxiRSeries2} which produces the sum
\begin{align}
\lim_{d \to 0}g_z&=\sum^\infty_{m=0}\frac{1}{(m(1-\lambda)+1)^3},\\
&=\sum^\infty_{k=0}\lambda^k\frac{(k+1)(k+2)}{2}\sum^k_{i=0}\frac{k!}{i!(k-i)!}(-1)^i\zeta(i+3).
\end{align}
where the second line expresses the first in terms of binomial sums of the Riemann zeta function $\zeta(z)$. In the infinite plane case ($\lambda=0$) only the first term remains equalling $\zeta(3)\approx1.2021$, which agrees with the result given by Cox and Brenner \cite{Cox_1967}. Interestingly, the increase in drag by the axisymmetric rotational wall effect from an an infinite plane is limited to less than just 20.3\%. As explored in other sections, axisymmetric rotation is the only kind of motion where the wall effect does not become singular in the small clearance limit.

\subsubsection{Axisymmetric Translational Wall Effect}
Axisymmetric translational wall effects of different sized spheres moving at the \textit{same} velocity were first evaluated by Stimson and Jeffery \cite{Stimson_1926}. Appendix \ref{appendix:AxiTraWE} outlines a modified version of their method where only the inner sphere moves along the $z$ axis with velocity $\nu$ while the outer sphere is stationary. Stimson and Jeffery identified a general series solution \eqref{eq_AxiTPsiwithC} to the axisymmetric Stokes equations, and related the $z$ component of the force to the series coefficients \eqref{eq_WEaxiTwithCoeffs}. We then determine the series coefficients using our different boundary conditions, giving the large series expression for $f_z$ shown in equation \eqref{eq_AxiTseries}.

By taking the Taylor series about $d=0$, we conjecture the singular nature of the small clearance limit of axisymmetric translating spheres to be
\begin{equation}\label{eq_AxiTSingular}
\frac{a}{(1-\lambda)^2 d}-\frac{1-7\lambda+\lambda^2}{5(1-\lambda)^3}\ln \frac{d}{a},
\end{equation}
which agrees with the result given by Cox and Brenner \cite{Cox_1967} in the infinite plane case $\lambda=0$.

\subsubsection{Asymmetric Wall Effects}
The asymmetrical wall effects, where the inner sphere rotates about, or translates along, an axis orthogonal to the line of displacement between the centres of the spheres, was first evaluated by Majumdar and O'Neill \cite{Majumdar_1969, ONeill_1970P1}. Similar to the axisymmetric cases, their method involves finding some series solutions to the equations of motion and then evaluating the coefficients to calculate the wall effects. Appendices \ref{appendix:AsyRotWE} and \ref{appendix:AsyTraWE} outline Majumdar and O'Neill's method, as well as introduce an improved technique for evaluating the series coefficients using forward differences, instead of backward differences. Majumdar and O'Neill related the wall effects to the series coefficients $E_n$ and $F_n$ by
\begin{align}
g_y&=\frac{\sqrt{2}}{4}\sinh^3\alpha\sum^\infty_{n=0}(2n+1-\coth\alpha)(E_n+F_n),\\
f_x^c&=\frac{\sqrt{2}}{3}\sinh^2\alpha\sum^\infty_{n=0}(E_n+F_n),\\
f_x&=\frac{\sqrt{2}}{3}\sinh\alpha\sum^\infty_{n=0}(E_n+F_n),
\end{align}
where $E_n$ and $F_n$ can be expressed in terms of $A_n$ and $B_n$ which are solved using our new forward differences method. Majumdar and O'Neill also provided expressions for the singular terms in the low clearance limit,
\begin{align}
g_y&=-\frac{2}{5}\frac{1}{1-\lambda}\ln\frac{d}{a}+\dots,\\
f_x&=-\frac{4}{15}\frac{2-\lambda+2\lambda^2}{(1-\lambda)^3}\ln\frac{d}{a}+\dots,\\
f_x^c&=-\frac{2}{15}\frac{4\lambda-1}{(1-\lambda)^2}\ln\frac{d}{a}+\dots.
\end{align}

\section{Machine Learning}\label{sec_MachineLearning}
The analytical solutions for the wall effects presented here are all in infinite series form, most of which present quite large expressions which are tedious to practically evaluate on a computer. The convergence of these series depend on $d/a$ and $a/b$ and for highly eccentric configurations can be quite slow, requiring hundreds of terms and high precision computation to be evaluated numerically. Therefore, it is useful to have a well established model that can replicate the wall effects much more efficiently and conveniently. By the universal approximation theorem \cite{Cybenko_1989}, a finite artificial neural network should be able to model these wall effects arbitrarily well. This section outlines the training and performance of such a model using data evaluated from the series solutions.

\subsection{Model}
The model should be able to compute the five dimensionless wall effects $f_x$, $f_z$, $g_y$, $g_z$ and $f_x^c$ from $d/a$ and $\lambda=a/b$ over as large a domain as possible. Of greatest importance, is the ability to evaluate the full dependence on the minimum clearance distance ($d$) for fixed radii $0<d\leqslant b-a$, as well as the transition behaviour between an infinite plane boundary ($\lambda=0$) and a finite spherical wall.

\subsubsection{Model Representation}
In the small clearance limit ($d/a\rightarrow0$) all of the wall effects (except $g_z$) become singular. Similarly, as the two spheres approach the same radii ($\lambda\rightarrow1$) all five effects also tend to infinity. The singular nature of the wall effects can be directly incorporated into the model since analytical expressions for the singular terms in both limits are known. Therefore, the neural network needs only learn the non-singular behaviour of the wall effects. The model ($\mathcal{W}$) is, therefore, comprised of the network ($\mathcal{N}$) which is then scaled by the concentric wall effects ($\mathcal{C}$), which accounts for the $\lambda\rightarrow1$ singularities, and added to modified low clearance singular terms ($\mathcal{S}$), which accounts for the $d/a\rightarrow0$ singularities,
\begin{equation}\label{eq_model}
\mathcal{W}(d/a,\lambda)=\mathcal{N}(\mathbf{x})\circ\mathcal{C}+\frac{\mathcal{S}}{1+(d/a)^2}.
\end{equation}
$\mathcal{W}$ is a vector of the dimensionless wall effects
\begin{equation}
\mathcal{W}=
\begin{bmatrix}
g_y & f_x^c & f_x & f_z & g_z
\end{bmatrix}^T,
\end{equation}
$\mathcal{C}$ represents a vector of the concentric wall effects given by equations \ref{eq_concentric_rotation} and \ref{eq_concentric_translation}
\begin{equation}
\mathcal{C}=
\begin{bmatrix}
g_{con} & f_{con} & f_{con} & f_{con} & g_{con}
\end{bmatrix}^T,
\end{equation}
$\circ$ represents the Hadamard product (element-wise multiplication) between $\mathcal{C}$ and $\mathcal{N}(\mathbf{x})$ (which is the neural network output) and $\mathcal{S}$ denotes a vector containing the corresponding singular terms given by equations (\ref{eq_singular_gy}), (\ref{eq_singular_fxc}), (\ref{eq_singular_fx}) and (\ref{eq_AxiTSingular}) respectively, and $0$ for the corresponding $g_z$ component. In the model this singular part is scaled down by $1+(d/a)^2$ so that the logarithmic terms do not diverge for large $d/a$. This is especially important for small $\lambda$ where the domain includes large values of $d/a$.

\subsubsection{Artificial Neural Network}
The final network chosen is a fully connected feed forward network with 2 inputs, 5 outputs and 50 nodes in the hidden layer. The network structure should be chosen to balance computation time with performance. We found that the network performance increased with the number hidden units, while remaining mostly invariant with the number of hidden layers. 50 nodes in a single hidden layer seemed to be enough to accurately fit the data while still being able to quickly compute the output. The inputs are normalised between $-1$ and 1 by
\begin{equation}
\mathbf{x}=\left[\frac{d/a-1}{d/a+1},2\lambda-1\right]^T.
\end{equation}

The hidden layer utilises a sigmoidal activation function defined by
\begin{equation}
\sigma(x)=\frac{2}{1+e^{-2x}}-1,
\end{equation}
while the output layer's activation function is linear.

Mathematically, the network is computed by
\begin{equation}
\mathcal{N}(\mathbf{x})=B2+W2\times\sigma(B1+W1\times\mathbf{x}),
\end{equation}
where $B1$ and $B2$ are column vectors containing the biases of each layer, $W1$ and $W2$ are matrices containing the weights of each layer, $\times$ represents matrix multiplication and $\sigma$ is applied component-wise. For reference, table \ref{table_net_coefficients} tabulates the trained values of these biases and weights to 8 significant figures.

\subsection{Data Evaluation and Network Training}
To train the network, training data was generated from the analytical results. $g_z$ and $f_z$ were calculated using equations (\ref{eq_AxiRSeries3}) and (\ref{eq_AxiTseries}). $g_y$, $f_x^c$ and $f_x$ were calculated using equations (\ref{eq_AsymWESeries1}), (\ref{eq_AsymWESeries2}) and (\ref{eq_AsymWESeries3}) with coefficients evaluated using section \ref{sec_recursive} methods. The truncation condition for each series was when the relative change in the finite sum by adding at least 10\% more terms was less than the desired precision ($10^{-16}$). The series expressions were evaluated using Mathematica software using a precision of 220. For most cases, this precision was much higher than necessary. However, to satisfy the truncation condition when $d/a$ was close to zero required hundreds of terms in the series, and using such a high precision was required when computing the asymmetric wall effects.

The training and validation data formed a random 70\% and 30\% split over a uniform $101\times91$ grid of $\frac{d}{b-a}\times\lambda$ over the domain
\begin{align}\label{eq_training_domain}
0.001\leqslant&\frac{d}{b-a}\leqslant0.999, & 0.05\leqslant&\lambda\leqslant0.95,
\end{align}
while an additional 2000 random points across the same domain formed the testing data. The network was trained in MATLAB using Levenberg--Marquardt backpropagation (trainlm) until the mean-squared error of the validation data stopped decreasing for 100 epochs. See Supplemental Material at [URL will be inserted by publisher] for the raw training data, testing data and a MATLAB implementation of the final model.

\subsection{Model Error}
After training the network, the performance of the full model, given by equation (\ref{eq_model}), over the training domain, equation (\ref{eq_training_domain}), can be quantified by the relative error between the model output and the random testing data.

\subsubsection{Training Region Performance}
\begin{figure}
\centering
\includegraphics*[width=\linewidth]{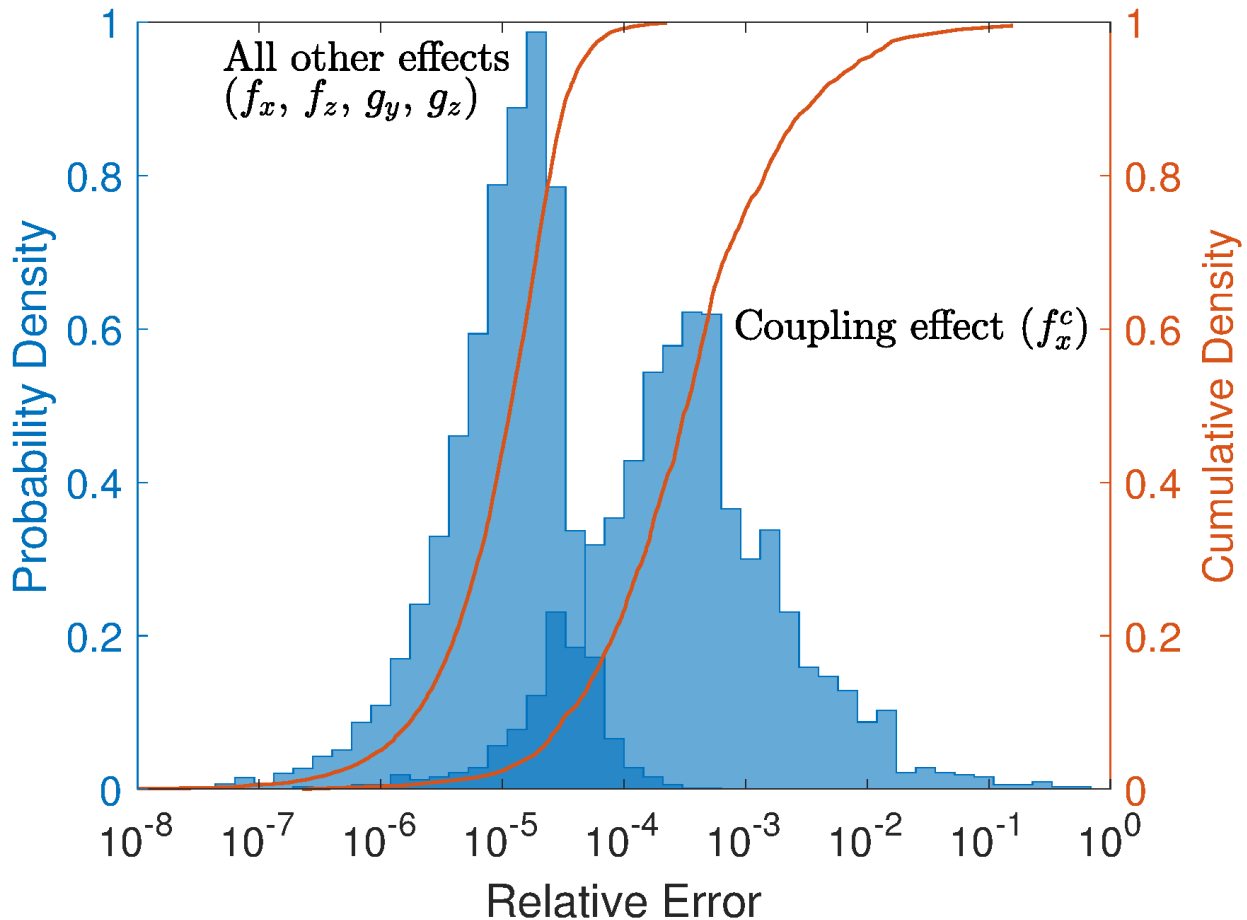}
\caption{Histograms of the relative error between model outputs and testing data. The coupling wall effect $f_x^c$ is kept seperate because of its larger relative errors. The solid red lines represent the cumulative densities showing the proportion of points less than the given relative error.}\label{fig_error_hist}
\end{figure}

Histograms of the relative errors are plotted in figure \ref{fig_error_hist}. These demonstrate two sets of behaviours with the model errors. The non-coupling wall effects $f_x$, $f_z$, $g_y$ and $g_z$ all exhibit similar relative errors, probably because they are all defined such that they are bounded by $\geqslant1$. The coupling effect $f_x^c$, however, tends to zero in the concentric limit and also decreases in magnitude in the $\lambda\rightarrow0$ limit. Therefore, the relative error in $f_x^c$ diverges, even for small absolute errors.

\begin{figure}
\centering
\includegraphics*[width=\linewidth]{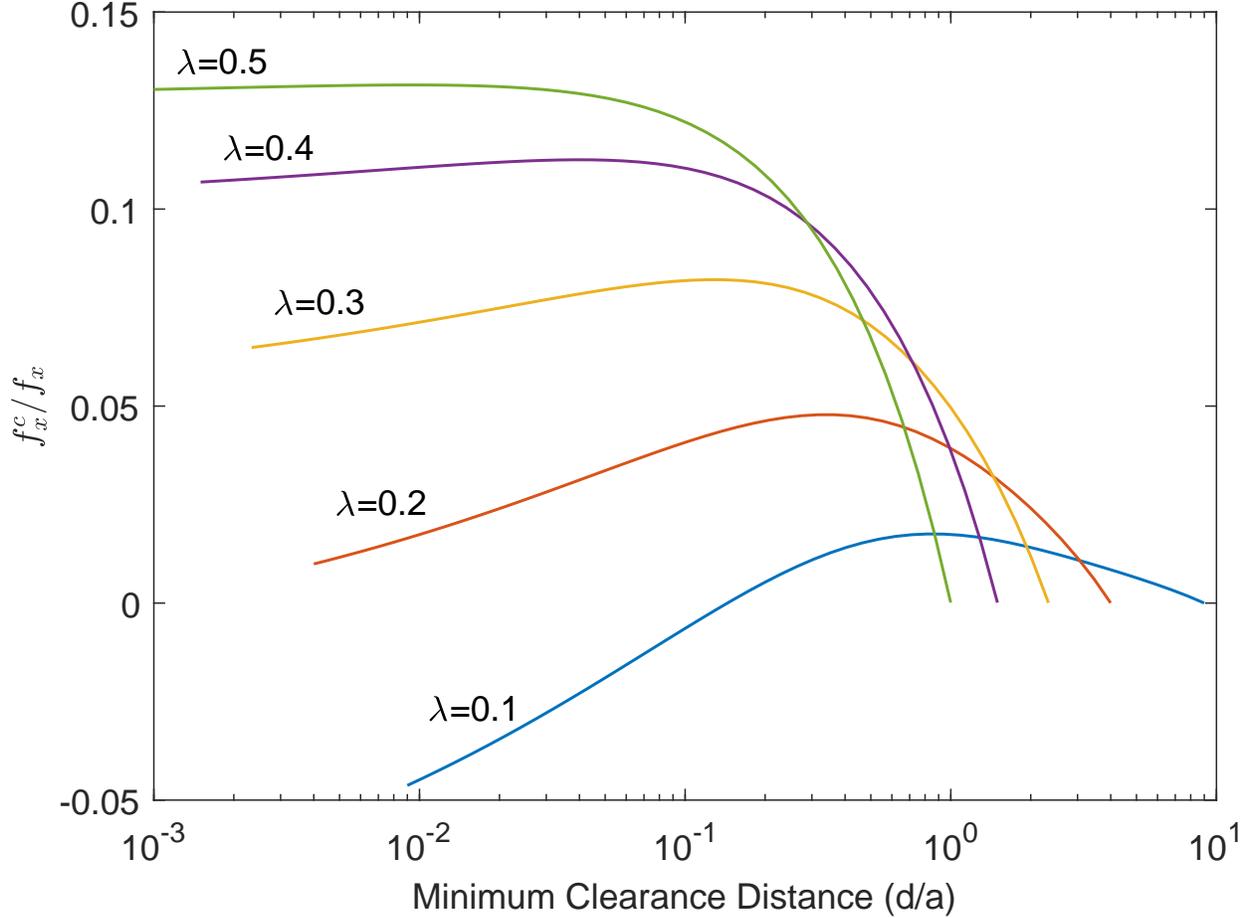}
\caption{The ratio of the asymmetric rotation-translation coupling force and the asymmetric translation is very small. It approaches zero in the concentric limit and decreases as $\lambda$ decreases.}\label{fig_coupling}
\end{figure}

In practice, large relative errors in coupling are less important when the other wall effects are much more significant. Figure \ref{fig_coupling} plots the ratio of the coupling wall effect $f_x^c$ with the corresponding asymmetrical translational wall effect $f_x$. The ratio tends to zero in the concentric limit and becomes smaller over a larger region as the outer sphere radius grows $\lambda\rightarrow0$. This demonstrates that the regions with higher relative error in the coupling wall effect, are the same regions where any wall effect from asymmetric translation or rotation would dominate.

Separating the coupling wall effect from the rest, the median relative error over the domain of training and validation data is $1.2\times10^{-5}$ and the maximum value is $5.1\times10^{-4}$. The median relative error of $f_x^c$ is $3.5\times10^{-4}$.

\begin{figure}
\centering
\includegraphics*[width=\linewidth]{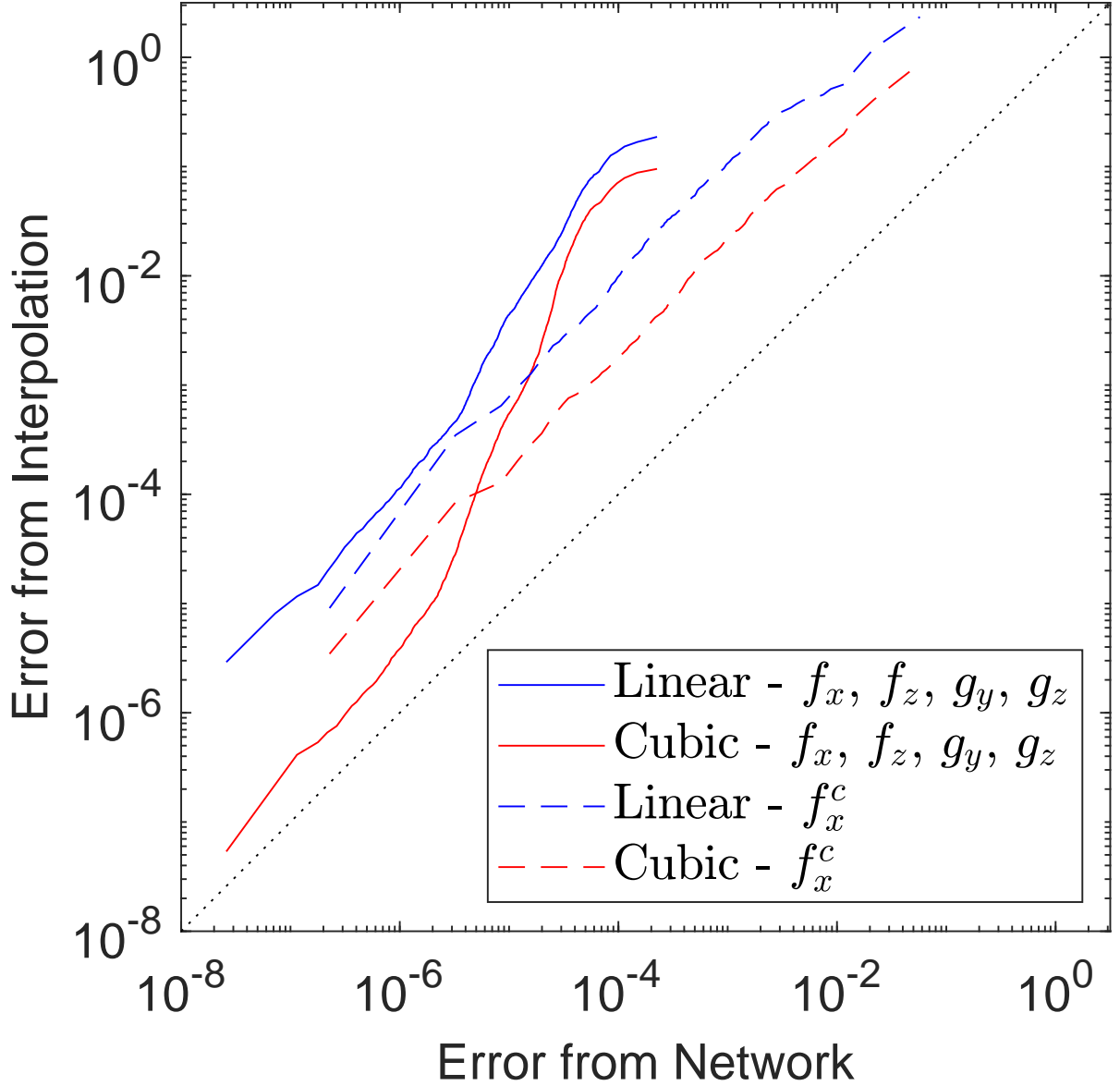}
\caption{A Q--Q plot comparing the performance of the model on the testing data when using the network and when using interpolation. The blue lines represent distributions of errors from linear interpolations over $11\times10$ grids of $\frac{d}{b-a}\times\lambda$. The red lines are corresponding results from cubic interpolations. The network outperforms the interpolation methods in all cases.}\label{fig_qqplot}
\end{figure}

Within the training domain, the model serves as an efficient system to interpolate between grid points, so it is worth comparing its performance to other interpolation techniques that use a comparable number of parameters. The network contains 405 weights and biases so choosing every tenth point in each dimension of the $101\times91$ grid gives 110 points for each of the 5 wall effects. This results in a total of 550 parameters, just a little more than the network. The model performs worse when applying linear or cubic interpolations over this grid instead of the network. Figure \ref{fig_qqplot} is a quantile--quantile (Q--Q) plot comparing the error distributions of the network performance (as shown in figure \ref{fig_error_hist}) with corresponding error distributions when applying linear and cubic interpolations. Evidently the network outperforms both forms of interpolation.

\subsubsection{Infinite Plan Extrapolation}
One of the goals of the model is to be able to model the transition behaviour between the eccentric sphere wall effects and the infinite plane wall effects. To test this, we check the relative error of the model when $\lambda=0$. The axisymmetric wall effects $g_z$ and $f_z$ could be evaluated using the same expressions but with $\beta=0$. The method for evaluating the asymmetric series coefficients becomes untenable in the infinite plane limit, so the approximations from Chaoui and Feuillebois \cite{Chaoui} for the asymmetric infinite plane wall effects were used instead.

\begin{figure}
\centering
\includegraphics*[width=\linewidth]{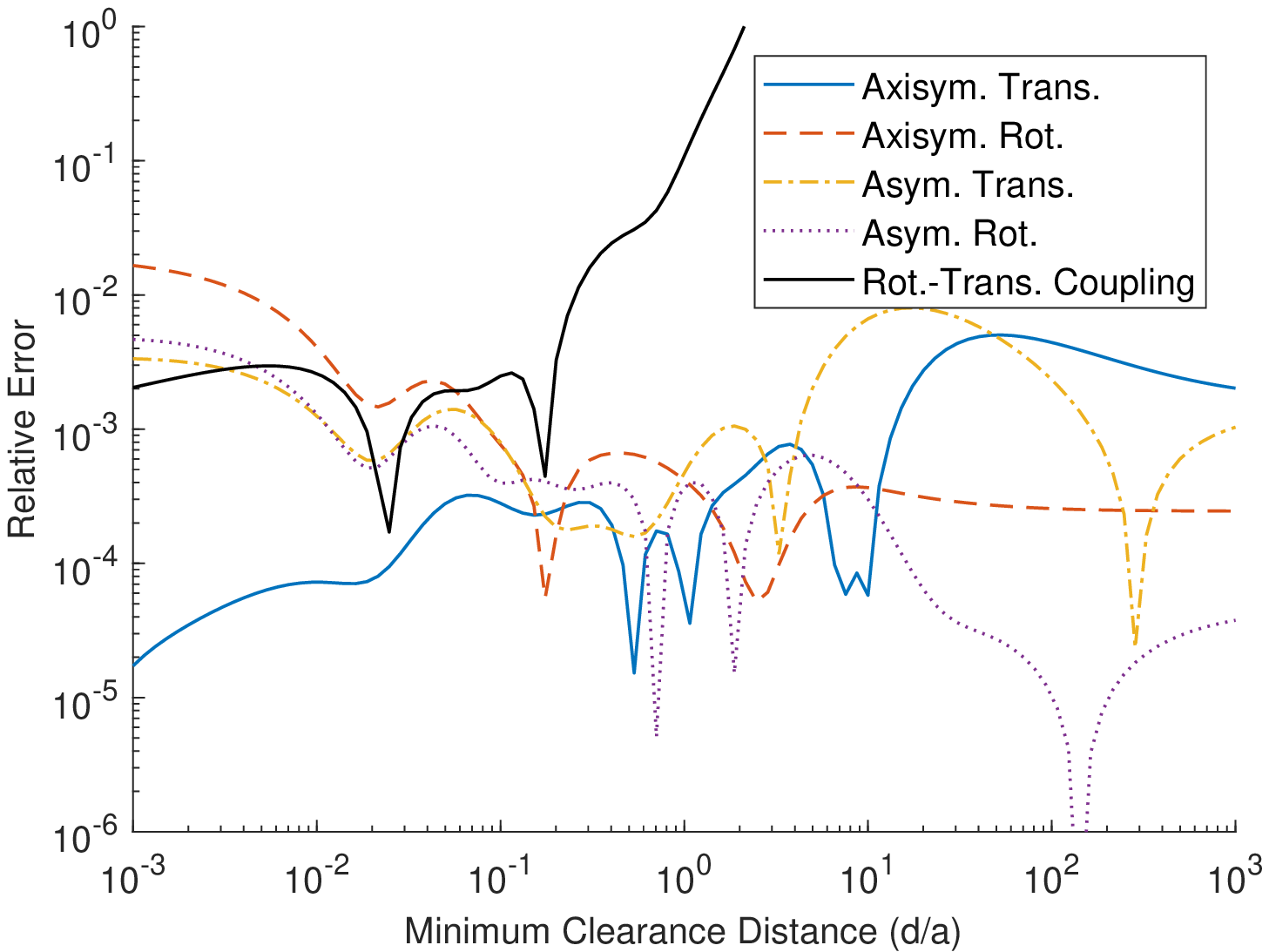}
\caption{The relative error of the model when extrapolating from the training region to calculate infinite plane wall effects ($\lambda=0$).}\label{fig_plane_wall_error}
\end{figure}

Figure \ref{fig_plane_wall_error} plots the relative errors of the model outputs as a function of $d/a$. Although the network was not trained on infinite plane wall effects, it did successfully reproduce them with a median relative error from non-coupling values of $4.6\times10^{-4}$ and maximum $1.7\times10^{-2}$, and a median coupling relative error of $f_x^c$ $1.1\times10^{-1}$. Although the relative error in coupling is comparatively large, this only occurs when the effect tends to zero and is marginal compared to the other wall effects. When $d/a<0.2$ the coupling effect becomes more significant but the model successfully evaluates it to less than 0.3\% error.

\section{Conclusion}
Analytical methods for calculating the wall effects of eccentrically positioned spheres have been improved and high precision evaluation of these effects over a discrete domain was able to generate data that could be used to train an artificial neural network to model the dimensionless forces and torques acting on the inner sphere. Within the training domain the model performed excellently on 2000 random test points with relative errors generally around 0.001\% error. The model successfully extrapolated to model the wall effects of an infinite plane on a sphere to less than 2\% error but generally around 0.05\%.

The success of the trained model using a relatively small network means that arbitrary motion of a sphere moving within another sphere can be efficiently modelled using easy-to-implement code. The model should be applicable for both experimental comparison as well as simulated dynamics which require small errors and high computational efficiency.

\begin{acknowledgments}
This research was supported under Australian Research Council's
Discovery Projects funding scheme (project number DP140100753) as well as an Australian Government Research Training Program Scholarship.
\end{acknowledgments}

\appendix

\section{Axisymmetric Rotational Wall Effect}\label{appendix:AxiRotWE}
\subsection{Series Solution}\label{sec_AxiRSeriesSolution}
The symmetry of the case where the inner sphere rotates axisymmetrically makes the mathematics comparatively simple. In this case only the rotational fluid velocity component, $v$, is non-zero and governed by a single equation,

\begin{equation} \label{eq_axisymm_motion}
\nabla^2v-\frac{v}{r^2}=0,
\end{equation}

\noindent while $u=w=p=0$ within the whole domain satisfies both the boundary conditions and equations of motion.

This problem has been previously solved analytically by Jeffery\cite{Jeffery1915} where he used a series solution to solve the equation of motion and evaluate the wall effect. Here we present an outline of a very similar derivation of the wall effect, noting that equation (\ref{eq_axisymm_motion}) has the bispherical series solution\cite{Jeffery_1912,Jeffery1915}
\begin{equation}
v=\sqrt{\cosh\varepsilon-\mu}\sum_{n=1}^\infty P_n^1(\mu)
\left[A_n\cosh\left(n+\frac{1}{2}\right)\varepsilon+B_n\sinh\left(n+\frac{1}{2}\right)\varepsilon\right],
\end{equation}
where $P_n^1(\mu)$ are associated Legendre functions, $\mu=\cos\psi$, and $A_n$ and $B_n$ are free coefficients.

In this form the total force and torque acting on the particle can be found via the integrals \ref{eq_ForceTotal} and \ref{eq_TorqueTotal}. The integrals for all vector components vanish except for the $z$ component of the torque ($G_z$) which is given by the infinite series
\begin{equation}
G_z=4\pi\sqrt{2}c^2\eta\sum_{n=1}^\infty n(n+1)(A_n+B_n).
\end{equation}
The dimensionless wall effect $g_z$ is evaluated by dividing this torque by the torque acting on a rotating sphere in a free fluid, 
\begin{align}
g_z&=\frac{G_z}{-8\pi\eta\Omega a^3}=\frac{G_z}{-8\pi\eta\Omega(c\csch\alpha)^3}\\
g_z&=-\frac{\sinh^3\alpha}{\Omega c\sqrt{2}}\sum_{n=1}^\infty n(n+1)(A_n+B_n)\label{eq_WEaxiRwithCoeffs}
\end{align}

\subsection{Evaluation of Coefficients}
Since a general series solution is available, the problem of evaluating the wall effect is reduced to evaluating the series coefficients. The orthogonality of the associated Legendre functions allow the coefficients to be evaluated analytically by enforcing the boundary conditions as outlined in section \ref{sec_BoundaryConditions}

%\begin{align}
%A_n\cosh\left(n+\frac{1}{2}\right)\alpha+B_n\sinh\left(n+\frac{1}{2}\right)\alpha&=-2\sqrt{2}c\Omega e^{-\left(n+\frac{1}{2}\right)\alpha}\\
%A_n\cosh\left(n+\frac{1}{2}\right)\beta+B_n\sinh\left(n+\frac{1}{2}\right)\beta&=0
%\end{align}

\begin{align}
A_n&=2\sqrt{2}c\Omega e^{-(n+\frac{1}{2})\alpha}\frac{\sinh\left(n+\frac{1}{2}\right)\beta}{\sinh(n+\frac{1}{2})(\alpha-\beta)}\\
B_n&=-2\sqrt{2}c\Omega e^{-(n+\frac{1}{2})\alpha}\frac{\cosh\left(n+\frac{1}{2}\right)\beta}{\sinh(n+\frac{1}{2})(\alpha-\beta)}.
\end{align}
Substituting these coefficients into equation (\ref{eq_WEaxiRwithCoeffs}) gives the axisymmetric rotational wall effect in series form
\begin{equation}\label{eq_AxiRSeries1}
g_z=4\sinh^3\alpha\sum^\infty_{n=1}\frac{n(n+1)}{e^{(2n+1)\alpha}-e^{(2n+1)\beta}}
\end{equation}

\subsection{Concentric Limit}
For high enough number of terms, the series in equation (\ref{eq_AxiRSeries1}) converges at a rate of $e^{-2\alpha}$. This means the series converges fastest for large $\alpha$ which occurs when the spheres are close to concentric. In the concentric limit
\begin{equation}
\lim_{\chi \to 0} e^{-\alpha}=\lim_{\chi \to 0} e^{-\beta}=0,\qquad \lim_{\chi \to 0} e^{\beta-\alpha}=\frac{a}{b}=\lambda.
\end{equation}
Therefore, in the case of concentric spheres all terms in the series vanish except for the first term which gives the well known result
\begin{equation}\label{eq_concentric_rotation}
\lim_{\chi \to 0}g_z=g_{con}=\frac{1}{1-\lambda^3}.
\end{equation}

\subsection{Alternative Series Expression}
Jeffery \cite{Jeffery1915} also gave an alternative series form of equation (\ref{eq_AxiRSeries1}) which converges at a different rate. By expanding the denominator of the summand in equation (\ref{eq_AxiRSeries1}) as a geometric series, a double summation can be produced
\begin{equation}
g_z=4\sinh^3\alpha\sum^\infty_{n=1}\sum^\infty_{m=0}n(n+1)e^{-(2n+1)(\alpha+m(\alpha-\beta))}.
\end{equation}
Next the summation order is switched and then the sum over $n$ can be simplified into a closed form expression giving the final result as a single (but different) summation
\begin{equation}\label{eq_AxiRSeries2}
g_z=\sum^\infty_{m=0}\left(\frac{\sinh\alpha}{\sinh(\alpha+m(\alpha-\beta))}\right)^3.
\end{equation}
The rate of convergence of this second series form is different $e^{-3(\alpha-\beta)}$ which means that in some configurations (such as the infinite plane case with $\beta=0$) this series converges faster.

\subsection{Combined Series Form}
We have managed to merge these different forms into a new combined sum which converges faster than both
\begin{align}\label{eq_AxiRSeries3}
\begin{split}
g_z&=\sum^{M}_{m=0}\left(\frac{\sinh\alpha}{\sinh(\alpha+m(\alpha-\beta))}\right)^3\\
&+4\sinh^3\alpha\sum^\infty_{n=1}\frac{n(n+1)e^{-(M+1)(2n+1)(\alpha-\beta)}}{e^{(2n+1)\alpha}-e^{(2n+1)\beta}}.
\end{split}
\end{align}
The first sum has the same summand as in equation (\ref{eq_AxiRSeries2}) but is truncated after the $m=M$ term. The second sum is a modified version of the sum \ref{eq_AxiRSeries1}, except the presence of the additional exponential factor improves the rate of convergence to $e^{-2(M+1)(\alpha-\beta)-2\alpha}$. Essentially, each term present in the first series improves the rate of convergence of the second by a factor of $e^{-2(\alpha-\beta)}$.

\subsection{Small Clearance Limit}
Although Jeffery \cite{Jeffery1915} correctly identified that these axisymmetric rotational wall effects near a plane wall were marginal, he failed to produce any expression for the wall effects in the $d\rightarrow0$ limit. This can be achieved by taking this limit of the summand in equation (\ref{eq_AxiRSeries2}) which produces the sum
\begin{equation}\label{eq_AxiR_d0}
\lim_{d \to 0}g_z=\sum^\infty_{m=0}\frac{1}{(m(1-\lambda)+1)^3}.
\end{equation}

The rate of convergence of this sum is quite slow, especially as $\lambda$ increases towards 1. Therefore, it might be useful to have approximate closed form expressions for this sum. By taking the Taylor series of the summand in equation (\ref{eq_AxiR_d0}), the series coefficients can be expressed in terms of binomial sums of the Riemann zeta function $\zeta(z)$
\begin{equation}\label{eq_AxiR_d02}
\lim_{d \to 0}g_z=\sum^\infty_{k=0}\lambda^k\frac{(k+1)(k+2)}{2}\sum^k_{i=0}\frac{k!}{i!(k-i)!}(-1)^i\zeta(i+3).
\end{equation}
In the infinite plane case ($\lambda=0$) only the first term remains equalling $\zeta(3)\approx1.2021$, which agrees with the result given by Cox and Brenner \cite{Cox_1967}. Interestingly, the increase in drag by the axisymmetric rotational wall effect from an an infinite plane is limited to less than just 20.2\%. As will be further explored in later sections, axisymmetric rotation is the only kind of motion where the wall effect does not become singular in the small clearance limit.

A reasonable approximation of the limiting wall effect when $\lambda<1/3$ can be acheived by taking the first few terms of equation (\ref{eq_AxiR_d02})
\begin{equation}\label{eq_AxiRd01}
g_{\lambda\rightarrow0}=\zeta(3)+3\lambda[\zeta(3)-\zeta(4)]+6\lambda^2[\zeta(3)-2\zeta(4)+\zeta(5)]
\end{equation}
where $\zeta(4)\approx1.0823$ and $\zeta(5)\approx1.0369$.

\begin{figure}
\centering
\includegraphics*[width=\linewidth]{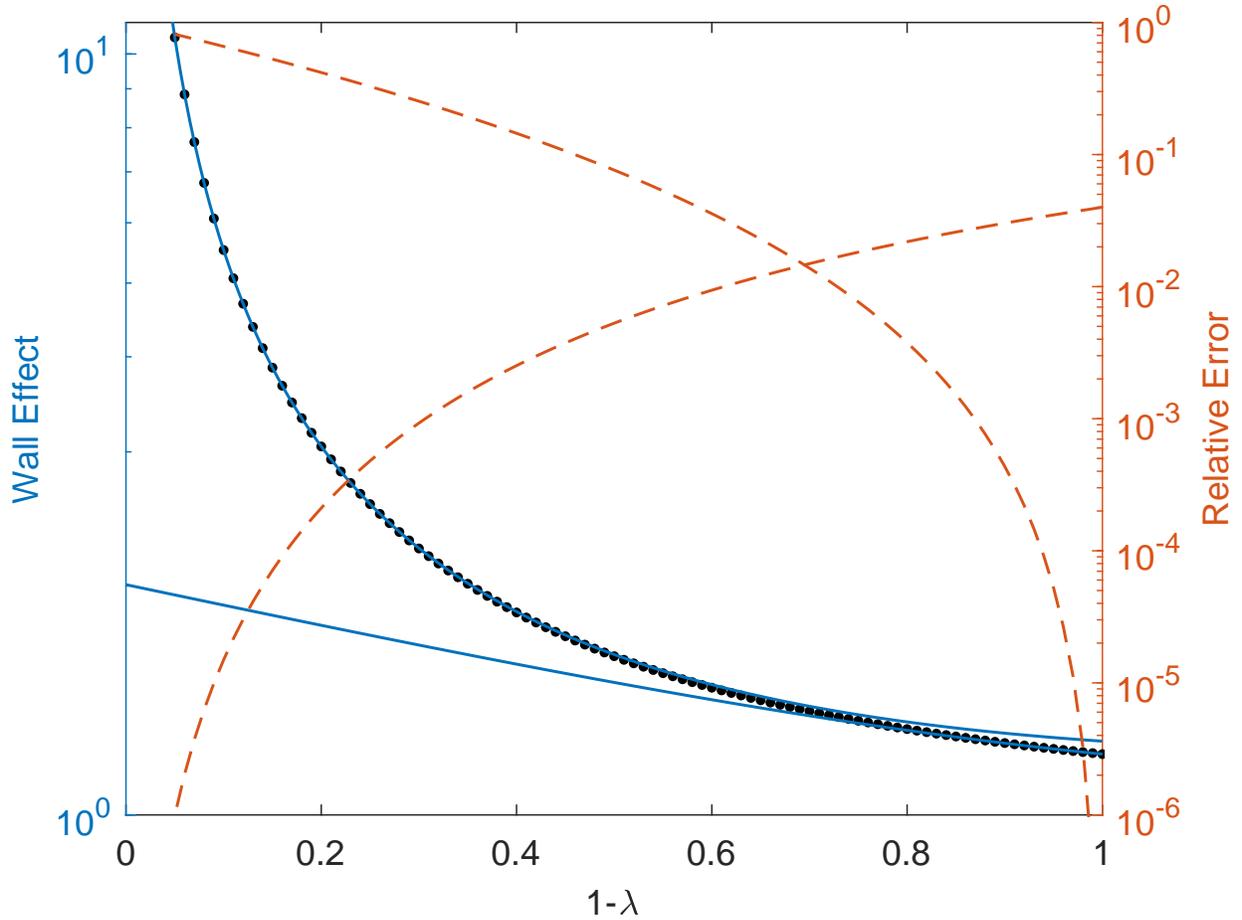}
\caption{A comparison between approximations for the axisymmetric rotational wall effect in the zero clearance limit. The black dots are the true values of the wall effects (calculated using many terms in sum \ref{eq_AxiR_d0}) while the solid lines are the $\lambda\rightarrow0$ and $\lambda\rightarrow1$ approximations given by equations \ref{eq_AxiRd01} and \ref{eq_AxiRd02} respectively. The relative errors of these approximations are plotted using the dashed lines and correspond to the right vertical axis.}\label{fig_axisymmetric_rotation_limit}
\end{figure}

Since the series converges slowest when $\lambda$ approaches 1, it would seem most useful to find a corresponding Taylor series about $\lambda=1$. However, applying the same method of expanding the summand in equation (\ref{eq_AxiR_d0}) yields divergent series. Resorting to empirical evaluation, the first few terms seem to be
\begin{equation}\label{eq_AxiRd02}
g_{\lambda\rightarrow1}=\frac{1}{2(1-\lambda)}+\frac{1}{2}+\frac{1}{4}(1-\lambda).
\end{equation}
Figure \ref{fig_axisymmetric_rotation_limit} compares $g_{\lambda\rightarrow0}$ and $g_{\lambda\rightarrow1}$ showing that their relative errors both tend to zero in their respective limits. If this empirical result is correct then the axisymmetric rotational wall effect is bounded by
\begin{equation}
\frac{1}{1-\lambda^3}\leqslant g_z<g_{\lambda\rightarrow1}<\frac{3}{2}\frac{1}{1-\lambda^3}.
\end{equation}

\section{Axisymmetric Translational Wall Effect}\label{appendix:AxiTraWE}
Axisymmetric translational wall effects of different sized spheres moving at the same velocity were first evaluated by Stimson and Jeffery \cite{Stimson_1926}. This section will outline a modified version of their method where only the inner sphere moves along the $z$ axis with velocity $\nu$ while the outer sphere is stationary.

\subsection{Stokes' Stream Function}
In this case of axisymmetric translation, the vertical and radial velocity components are non-zero while the rotational component is zero. Therefore, Stimson and Jeffery expressed the velocity components in terms of Stokes' stream function $\Psi$
\begin{equation}
u=\frac{1}{r}\pd{\Psi}{z},\qquad v=0,\qquad w=-\frac{1}{r}\pd{\Psi}{r}.
\end{equation}
Noting that all derivatives with respect to $\theta$ are zero (because of axisymmetry), they showed from the equations of motion (equations \ref{eq_equation_of_motion1} and \ref{eq_equation_of_motion3}) that the stream function must satisfy the linear partial differential equation
\begin{equation}
\Phi^4\Psi=0
\end{equation}
where $\Phi^2$ is the linear differential operator defined by
\begin{equation}
\Phi^2=r\pd{}{r}\left(\frac{1}{r}\pd{}{r}\right)+\pd[2]{}{z}
\end{equation}
thus reducing the problem down to solving a single equation of a single function $\Psi$.

\subsection{Series Solution}
This equation has a similar series solution in bispherical coordinates to the axisymmetric rotational case shown in section \ref{sec_AxiRSeriesSolution}, except there are now four sets of free coefficients $A_n$, $B_n$, $C_n$ and $D_n$ (not the same values as before) \cite{Stimson_1926,Jeffery_1912}
\begin{multline}\label{eq_AxiTPsiwithC}
\Psi=(\cosh\varepsilon-\mu)^{-\frac{1}{2}}\sum^\infty_{n=1}(P_{n-1}(\mu)-P_{n+1}(\mu))[\\
A_n\cosh\left(n-\frac{1}{2}\right)\varepsilon+B_n\sinh\left(n-\frac{1}{2}\right)\varepsilon\\
+C_n\cosh\left(n+\frac{3}{2}\right)\varepsilon+D_n\sinh\left(n+\frac{3}{2}\right)\varepsilon].
\end{multline}

Similar to the rotational case, evaluating the integrals given in equations (\ref{eq_ForceTotal}) and (\ref{eq_TorqueTotal}) for each term in the sum can give the total force and torque acting on the particle. The integrals for all vector components vanish except for the $z$ component of the force ($F_z$) which is given by the infinite series
\begin{equation}
F_z=\frac{2\pi\eta\sqrt{2}}{c}\sum_{n=1}^\infty (2n+1)(A_n+B_n+C_n+D_n).
\end{equation}
The dimensionless wall effect $f_z$ is evaluated by dividing this force by the corresponding force acting on a translating sphere in a free fluid, 
\begin{align}
f_z&=\frac{F_z}{-6\pi\eta\nu a}=\frac{F_z}{-6\pi\eta\nu c\csch\alpha}\\
f_z&=-\frac{\sqrt{2}\sinh\alpha}{3c^2\nu}\sum_{n=1}^\infty (2n+1)(A_n+B_n+C_n+D_n).\label{eq_WEaxiTwithCoeffs}
\end{align}

\subsection{Evaluation of Coefficients}
The problem of finding the axisymmetric rotational wall effects has now been reduced to evaluating the series coefficients. This is again achieved by enforcing the boundary conditions in equation (\ref{eq_TranslationalBCs}). Since we use different boundary conditions (outer sphere is stationary rather than translating), this is also the point where our calculation differs from Stimson and Jeffery's \cite{Stimson_1926}. The boundary conditions can be expressed in terms of the stream function by
\begin{align}
\text{at } \varepsilon&=\alpha &\pd{\Psi}{z}&=0, &\pd{\Psi}{r}&=-r\nu,\\
\text{at } \varepsilon&=\beta &\pd{\Psi}{z}&=0, &\pd{\Psi}{r}&=0.
\end{align}

Combining the four boundary conditions with the series solution for $\Psi$ given by equation (\ref{eq_AxiTPsiwithC}) and then exploiting the orthogonality of the Legendre polynomials gives a system of simultaneous equations for $A_n$, $B_n$, $C_n$ and $D_n$. For brevity this system is omitted here but a close version can be seen by equation (26) in \cite{Stimson_1926}. Upon solving the system and substituting back into equation (\ref{eq_WEaxiTwithCoeffs}) yields the rather large expression for the wall effect
\begin{widetext}
\begin{equation}\label{eq_AxiTseries}
f_z=\sinh\alpha\sum^\infty_{n=1}\frac{4n(n+1)}{3(2n-1)(2n+3)}\frac{e^{-(2n+1)\beta}(f(\alpha,n)+(4n+2)\sinh2\alpha)-e^{-(2n+1)\alpha}(f(\beta,n)+(4n+2)\sinh2\beta)}{4\cosh(2n+1)(\alpha-\beta)-f(\alpha-\beta,n)},
\end{equation}
\end{widetext}

\begin{equation}
\text{where } f(\varepsilon,n)=4+(2n+1)^2(\cosh2\varepsilon-1).
\end{equation}

\subsection{Concentric Limit}
Although less obvious from the expression, like the rotational case, all except the first term vanish in the concentric limit. The first term becomes the well known translational wall effect for concentric spheres
\begin{equation}\label{eq_concentric_translation}
\lim_{\chi \to 0}f_z=f_{con}=\frac{4(1-\lambda^5)}{(1-\lambda)^4(4+7\lambda+4\lambda^2)}.
\end{equation}

\subsection{Small Clearance Limit}
Similar to the rotational case, we can try to obtain limiting expressions for the axisymmetric translational wall effect in the small clearance limit. Taking the Taylor series of the summand in equation (\ref{eq_AxiTseries}) about $d=0$ suggests $1/d$ dependence for small clearances
\begin{align}
\begin{split}
f_z&=\sum^\infty_{n=1}\frac{32n(n+1)}{(2n-1)^2(2n+1)(2n+3)^2}\frac{a}{(1-\lambda)^2 d}+\mathcal{O}(d^0)\\
f_z&=\frac{a}{(1-\lambda)^2 d}+\sum^\infty_{n=1}\text{harmonic term}+\mathcal{O}(\sqrt{d}).
\end{split}
\end{align}
The sum of the $1/d$ term evaluates to a relatively simple closed form. However, the coefficients of the constant term and the following terms of powers of $\sqrt{d}$ form divergent series. Part of the constant term is related to the harmonic series $\sum 1/n$ which suggests the existence of a $\ln d$ singularity. Motivated by the $\sqrt{d}$ powers of later terms, we conjecture that the logarithmic term is proportional to \textit{half} the coefficient of the harmonic-like series. Therefore, the singular nature of the small clearance limit of axisymmetric translating spheres is conjectured to be
\begin{equation}\label{eq_AxiTSingular2}
\frac{a}{(1-\lambda)^2 d}-\frac{1-7\lambda+\lambda^2}{5(1-\lambda)^3}\ln \frac{d}{a}.
\end{equation}
Note that in the infinite plane case $\lambda=0$ this agrees with the result given by Cox and Brenner \cite{Cox_1967}.

\begin{figure}
\centering
\includegraphics*[width=\linewidth]{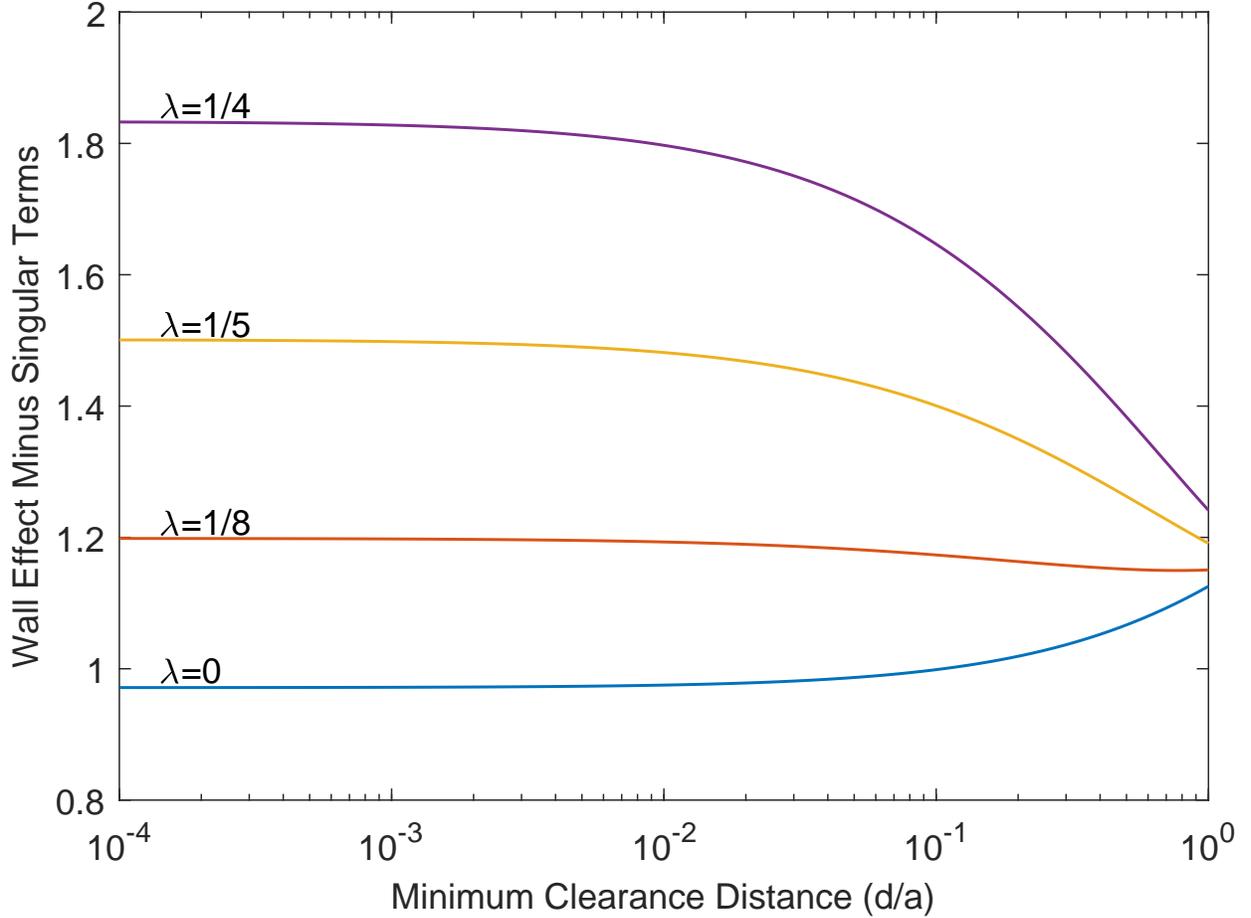}
\caption{The axisymmetric translational wall effects (equation (\ref{eq_AxiTseries})) subtract the conjectured singular terms (equation (\ref{eq_AxiTSingular2})) appear to converge towards finite values in the zero clearance limit.}\label{fig_axisymmetric_translation_limit}
\end{figure}

\section{Asymmetric Rotational Wall Effect} \label{appendix:AsyRotWE}
The asymmetrical rotational wall effect, where the inner sphere rotates about an axis orthogonal to the line of displacement between the centres of the spheres, was first evaluated by Majumdar \cite{Majumdar_1969} and then further refined in collaboration with O'Neill \cite{ONeill_1970P1}. Similar to the axisymmetric cases, the method here involves finding some series solutions to the equations of motion and then evaluating the coefficients to calculate the wall effects. This section will outline Majumdar and O'Neill's method, and introduce an improved method for evaluating the series coefficients.

\subsection{Dimensionality Reduced Stokes Equations}
Although the rotation of the inner particle is asymmetric, Majumdar was still able to eliminate $\theta$ from the equations of motion (\ref{eq_equation_of_motion1}-\ref{eq_equation_of_motion5}) and boundary conditions (\ref{eq_RotationalBCs}) by using the following variable transformation
\begin{align}
u&=1/2~\Omega(rQ_1+cU_2+cU_0)\cos\theta\label{eq_asymtransform1}\\
v&=1/2~\Omega(cU_2-cU_0)\sin\theta\\
w&=1/2~\Omega(zQ_1+2cw_1)\cos\theta\\
p&=\eta\Omega Q_1 \cos\theta\label{eq_asymtransform4}
\end{align}
where $U_0$, $U_2$, $w_1$ and $Q_1$ are dimensionless functions independent of $\theta$. The equations of motion reduce to
\begin{equation}\label{eq_TransformedEquationsofMotion}
L_0^2U_0=L_2^2U_2=L_1^2w_1=L_1^2Q_1=0
\end{equation}
where $L_m^2$ is a class of linear differential operators defined by
\begin{equation}
L_m^2=\frac{\partial^2}{\partial r^2}+\frac{1}{r}\frac{\partial}{\partial r}-\frac{m^2}{r^2}+\frac{\partial^2}{\partial z^2},
\end{equation}
and the continuity equation transforms to
\begin{equation}\label{eq_TransformedContinuity}
\left[3+r\frac{\partial}{\partial r}+z\frac{\partial}{\partial z}\right]Q_1
+c\left[\frac{\partial U_0}{\partial r}+\left(\frac{\partial}{\partial r}+\frac{2}{r}\right)U_2+2\frac{\partial w_1}{\partial z}\right]=0.
\end{equation}

%In both the asymmetric cases the boundary conditions involve factors of $\cos\theta$ and $\sin\theta$. Therefore, we are motivated to seek solutions of the form
%
%\begin{equation}
%\begin{bmatrix}
%u\\
%v\\
%w\\
%p
%\end{bmatrix}
%=\Omega
%\begin{bmatrix}
%Q(r,z)\cos\theta\\
%S(r,z)\sin\theta\\
%T(r,z)\cos\theta\\
%\eta P(r,z)\cos\theta
%\end{bmatrix}.
%\end{equation}
%
%Applying this variable transformation does reduce the dimensionality of the equations of motion to
%
%\begin{align}
%\nabla^2Q-\frac{2Q}{r^2}-\frac{2S}{r^2}&=\pd{P}{r},\\
%\nabla^2S-\frac{2S}{r^2}-\frac{2Q}{r^2}&=-\frac{P}{r},\\
%\nabla^2T-\frac{T}{r^2}&=\pd{P}{z},\\
%\pd{Q}{r}+\frac{Q}{r}+\frac{S}{r}+\pd{T}{z}&=0,\\
%\nabla^2P-\frac{P}{r^2}&=0.
%\end{align}

\subsection{Series Solution}
Similar to the axisymmetric cases, the transformed equations of motion \ref{eq_TransformedEquationsofMotion} have series solutions in bispherical coordinates \cite{Jeffery_1912,Majumdar_1969,ONeill_1970P1}
\begin{equation}\label{eq_AsymSeries1}
w_1=(\cosh\varepsilon-\mu)^{1/2}\sum\limits_{n=1}^\infty P_n^1(\mu)[
A_n\cosh(n+1/2)\varepsilon+B_n\sinh(n+1/2)\varepsilon],
\end{equation}
\begin{equation}
Q_1=(\cosh\varepsilon-\mu)^{1/2}\sum\limits_{n=1}^\infty P_n^1(\mu)[
C_n\cosh(n+1/2)\varepsilon+D_n\sinh(n+1/2)\varepsilon],
\end{equation}
\begin{equation}
U_0=(\cosh\varepsilon-\mu)^{1/2}\sum\limits_{n=0}^\infty P_n(\mu)[
E_n\cosh(n+1/2)\varepsilon+F_n\sinh(n+1/2)\varepsilon],
\end{equation}
\begin{equation}
U_2=(\cosh\varepsilon-\mu)^{1/2}\sum\limits_{n=2}^\infty P_n^2(\mu)[
G_n\cosh(n+1/2)\varepsilon+H_n\sinh(n+1/2)\varepsilon].
\end{equation}
However, this time there are eight sets of coefficients $A_n$, $B_n$, $C_n$, $D_n$, $E_n$, $F_n$, $G_n$ and $H_n$.

Using these series solutions Majumdar and O'Neill \cite{ONeill_1970P1} managed to relate the wall effects to just the $E_n$ and $F_n$ coefficients by 
\begin{align}
g_y&=\frac{\sqrt{2}}{4}\sinh^3\alpha\sum^\infty_{n=0}(2n+1-\coth\alpha)(E_n+F_n),\label{eq_AsymWESeries1}\\
f_x^c&=\frac{\sqrt{2}}{3}\sinh^2\alpha\sum^\infty_{n=0}(E_n+F_n).\label{eq_AsymWESeries2}
\end{align}

\subsection{Recursive Coefficient System}\label{sec_recursive}
The eight sets of coefficients are determined by both the boundary conditions given by equation (\ref{eq_RotationalBCs}) and the continuity equation (\ref{eq_TransformedContinuity}). Through these constraints Majumdar \cite{Majumdar_1969,ONeill_1970P1} and O'Neill \cite{ONeill_1970P1} were able to express all other coefficients in terms of $A_n$ and $B_n$, and relate $A_n$ and $B_n$ through two sets of simultaneous recursive equations
\begin{align}\label{eq_AsymRecursive}
\mathcal{R}^1*(A_n,B_n)&=i_n, & \mathcal{R}^2*(A_n,B_n)&=j_n
\end{align}
where $\mathcal{R}^i*(A_n,B_n)$ is defined by
\begin{equation}
\mathcal{R}^i*(A_n,B_n)=a^i_n A_{n-1}+b^i_{n}B_{n-1}
+c^i_{n}A_{n}+d^i_n B_{n}+e^i_n A_{n+1}+f^i_n B_{n+1}.
\end{equation}
For brevity the expressions for $a^i_n$ - $f^i_n$, $i_n$ and $j_n$ are omitted here but are given by Majumdar \cite{Majumdar_1969} in equations 39 and 40.

For any given values of $\alpha$ and $\beta$, Majumdar and O'Neill \cite{ONeill_1970P1} solve the system numerically by truncating the system at sufficiently high order and solving the finite system using a Gauss--Seidel method. In the case of eccentric spheres, where one sphere is enclosed by the other, equations (\ref{eq_AsymRecursive}) approach dependence for large $n$ and so the system becomes singular if truncated at too high order. This poses a problem, especially since the most important dominant lower order coefficients are evaluated through backward difference from the point of truncation.

Therefore, for high precision calculations of these coefficients it would seem much better to somehow evaluate $A_1$ and $B_1$ and use forward difference to solve subsequent values. In the correspoinding infinite plane problem there is only a single recursive equation which O'Neill and Bhatt \cite{ONEILL_BHATT_1991} and Chaoui \cite{Chaoui} solved by transforming $A_n$ into a combination of two other coefficients which are related to $A_n$ by $A_1$. This allowed them to calculate the transformed coefficients using forward difference and then estimate $A_1$ by their limiting behaviour.

Motivated by this technique, we transform $A_n$ and $B_n$ into
\begin{align}\label{eq_coeffs_transform}
\begin{split}
A_n&=T_n+A_1 U_n+B_1 V_n\\
B_n&=W_n+A_1 X_n+B_1 Y_n
\end{split}
\end{align}
where $T_n$, $U_n$, $V_n$, $W_n$, $X_n$ and $Y_n$ are new transformed coefficients satisfying
\begin{align}
T_1=W_1=V_1=X_1&=0, & U_1=Y_1&=1,
\end{align}
and the following recursive equations
\begin{align}
&\mathcal{R}^1*(T_n,W_n)=i_n, & &\mathcal{R}^2*(T_n,W_n)=j_n,\\
&\mathcal{R}^1*(U_n,X_n)=0, & &\mathcal{R}^2*(U_n,X_n)=0,\\
&\mathcal{R}^1*(V_n,Y_n)=0, & &\mathcal{R}^2*(V_n,Y_n)=0.
\end{align}
It is easy to show that this transformation is consistent with the original recursive equations \ref{eq_AsymRecursive} except the new coefficients can easily be evaluated using forward differences.

The last step involves evaluating $A_1$ and $B_1$, which can be done through the limiting behaviour of the transformed coefficients. In particular, equation (\ref{eq_coeffs_transform}) can be inverted to
\begin{align}
\begin{split}
A_1&=\frac{(W_n-B_n)V_n-(T_n-A_n)Y_n}{U_nY_n-X_nV_n},\\
B_1&=\frac{(W_n-B_n)U_n-(T_n-A_n)X_n}{V_nX_n-Y_nU_n}.
\end{split}
\end{align}

Assuming the original series solution in equation (\ref{eq_AsymSeries1}) converges, the coefficients $A_n$ and $B_n$ must tend to zero as $n\rightarrow\infty$. Therefore, we are motivated to define a sequence of approximate values of $A_1$ and $B_1$ as
\begin{align}
A_1^n&=\frac{W_nV_n-T_nY_n}{U_nY_n-X_nV_n}, & B_1^n&=\frac{W_nU_n-T_nX_n}{V_nX_n-Y_nU_n},
\end{align}
which should satisfy
\begin{align}
\lim_{n\rightarrow\infty}A_1^n&=A_1, & \lim_{n\rightarrow\infty}B_1^n&=B_1,
\end{align}
if $W_n$ and $T_n$ do not converge to 0. In practice we find that this is the case and that $A_1^n$ and $B_1^n$ converge like
\begin{equation}
A_1^n-A_1^{n-1}\propto B_1^n-B_1^{n-1}\propto ne^{-2\beta n}.
\end{equation}
This limiting behaviour can be extrapolated to infinity to improve the approximation for finite orders
\begin{align}\label{eq_AsymA1B1}
\begin{split}
A_1&\approx A_1^n+(A_1^n-A_1^{n-1})e^{-2\beta}\frac{1+n(1-e^{-2\beta})}{n(1-e^{-2\beta})^2},\\
B_1&\approx B_1^n+(B_1^n-B_1^{n-1})e^{-2\beta}\frac{1+n(1-e^{-2\beta})}{n(1-e^{-2\beta})^2}.
\end{split}
\end{align}

Therefore, using equation (\ref{eq_AsymA1B1}) to calculate $A_1$ and $B_1$, equation (\ref{eq_coeffs_transform}) can be used to evaluate $A_n$ and $B_n$.

\subsection{Small Clearance Limit}
Expressions for the singular terms of the asymmetrical wall effects are given by O'Neill and Majumdar \cite{ONeill_1970P2}
\begin{align}
g_y&=-\frac{2}{5}\frac{1}{1-\lambda}\ln\frac{d}{a}+\dots,\label{eq_singular_gy}\\
f_x&=-\frac{4}{15}\frac{2-\lambda+2\lambda^2}{(1-\lambda)^3}\ln\frac{d}{a}+\dots,\label{eq_singular_fx}\\
f_x^c&=-\frac{2}{15}\frac{4\lambda-1}{(1-\lambda)^2}\ln\frac{d}{a}+\dots.\label{eq_singular_fxc}
\end{align}

\section{Asymmetric Translational Wall Effect} \label{appendix:AsyTraWE}
The asymmetric translational wall effects are evaluated almost identically to the preceding asymmetric rotational wall effects with only a few minor differences. The equations of motion are again reduced to equation (\ref{eq_TransformedEquationsofMotion}) by applying the same variable transformation as equations \ref{eq_asymtransform1}--\ref{eq_asymtransform4} except with the replacement $\Omega\rightarrow\nu/c$. The same series solutions are utilised except the coefficients' values are different. The dimensionless force is given by
\begin{equation}\label{eq_AsymWESeries3}
f_x=\frac{\sqrt{2}}{3}\sinh\alpha\sum^\infty_{n=0}(E_n+F_n).
\end{equation}

The expressions for $i_n$ and $j_n$ are different in the translation case because the boundary conditions are different. However, with the exception of this difference, the coefficients $A_n$ and $B_n$ can be evaluated the same way as in the rotational case.

\squeezetable
\begin{table*}[!h]
\begin{tabular}{c|cc|ccccc}
$B1$ & \multicolumn{2}{c|}{$W1$} & \multicolumn{5}{c}{$W2^T$}\\
\hline
4.8113394 & -2.9231488 & -0.19127825 & -6.3213834 & -18.123169 & -12.355483 & 32.938574 & -1.9663136\\
4.5643461 & -10.860592 & -9.8976544 & -19.759477 & -10.955627 & -15.427646 & 19.596163 & -6.892128\\
-4.5063286 & 2.5794291 & -0.57787839 & -1.6816723 & -5.2511235 & -3.2960726 & 8.1682192 & -0.56582731\\
5.211532 & -0.89619242 & 3.3242802 & -0.12901484 & -0.58020697 & -0.33093263 & 0.41179354 & -0.065380236\\
2.6380771 & -3.3084402 & -4.3683345 & -4.8869387 & 0.68458723 & -0.082493243 & -1.8959001 & -0.79845435\\
12.991963 & -11.787911 & -24.2786 & 14.790883 & -4.810932 & -0.11552022 & 4.1409808 & 1.0302539\\
10.440263 & -21.973296 & -30.476054 & 14.07486 & -6.5362532 & -0.206136 & -0.59875748 & 0.80570816\\
-1.9661939 & -0.93824277 & -0.74110631 & 1.8272341 & -15.152262 & -4.4719601 & 16.482887 & 11.625042\\
1.6562525 & -0.508093 & 1.0002596 & -2.7220653 & -2.7714042 & -2.9031277 & 10.906282 & -0.87789713\\
1.5871351 & -0.59273394 & 0.8478989 & 3.630631 & 3.2871044 & 2.5964378 & -13.685592 & 1.0218652\\
-18.154178 & -5.5214697 & 10.740687 & 0.40812945 & -17.993253 & 0.042159326 & 9.2202538 & -7.3656119\\
-101.1157 & -87.529199 & 12.362803 & -3.5990025 & -2.83695 & 0.050704842 & 4.1705861 & 2.2782291\\
0.1394124 & -1.4397034 & 0.18357732 & -0.79059739 & -1.1726016 & 1.682904 & -5.5206363 & 0.11718537\\
-0.34850942 & -0.62701717 & -0.58818834 & -0.95215953 & 0.66840542 & -1.8153526 & 4.9081699 & -0.63673305\\
4.1411886 & -10.308195 & -9.1576609 & 7.1192099 & 3.3686357 & 6.4683209 & -15.307482 & 2.4936587\\
-0.079175964 & 1.2190827 & 0.083425231 & 4.999372 & 4.0872406 & 0.94409221 & -2.6496287 & 0.22328977\\
0.11709931 & 0.46571726 & -0.34610801 & -4.5114092 & 0.29811614 & 3.1590265 & 12.937801 & -1.3904952\\
3.1701055 & -0.23574607 & -3.4757205 & 0.73913844 & -0.39776158 & 0.022167735 & 0.28666821 & 0.035013077\\
-12.493281 & 13.457111 & 24.860202 & 45.88042 & -14.284513 & -0.3590089 & 12.601054 & 3.5595016\\
0.13945188 & -1.5153369 & 0.13135781 & 1.3255305 & 1.3980774 & -1.1914806 & 3.7415379 & -0.091354023\\
-5.509932 & 5.1259729 & 9.4554294 & 12.578025 & -2.4511914 & -0.18342767 & 4.6062157 & 1.712977\\
0.2407358 & -1.2733534 & -0.058980096 & 3.9380972 & 3.7897635 & 0.66840072 & -0.51027143 & 0.077155285\\
1.6609392 & 0.36350978 & -1.3092586 & 1.12103 & -0.63803385 & 0.36837789 & 1.1595936 & 0.29330014\\
-1.8788659 & -1.9290912 & 0.41248143 & 0.69266209 & 0.30858376 & -0.11972862 & -1.0364245 & -0.3576897\\
-0.38260522 & -0.14451115 & 0.18892011 & -9.1871081 & -7.8246746 & 24.299862 & 16.922558 & 3.0753937\\
-4.2833535 & 9.6026275 & 11.297067 & -2.5293361 & 2.0056763 & 0.15567844 & 4.5601494 & -0.42652554\\
-3.3792586 & -2.4819184 & 0.48181726 & 9.3875976 & 6.5688575 & -0.041558872 & -12.566227 & -5.9786109\\
6.7632512 & -2.9390105 & -8.3092822 & 29.295549 & -8.6337375 & -0.26950606 & 9.5683882 & 2.9499413\\
-3.3592869 & -2.0614463 & 0.3742545 & -26.594558 & -19.56772 & -0.72779693 & 34.762897 & 18.847575\\
0.35806924 & -1.7091651 & -1.1349919 & -0.1240636 & -0.10956647 & -0.14858281 & 0.41901725 & -0.091429445\\
-2.5564898 & -1.4286332 & 1.3590837 & -0.21646259 & -0.49888707 & 0.014728244 & 0.55719107 & -0.10686906\\
2.4520402 & -2.4781979 & -2.801879 & 0.59375222 & -1.28546 & 0.93238021 & -6.480926 & -1.2092875\\
20.383505 & 19.792641 & 0.5185137 & 3.8065898 & 3.8804101 & 0.083159283 & -5.8080313 & -4.6632913\\
0.16392206 & 0.05736511 & -0.49034398 & 2.4749204 & -2.8603028 & 8.9343512 & -11.453807 & 3.3752432\\
20.773334 & 20.332416 & 0.30276654 & 3.3112701 & 3.3967786 & 0.1242549 & -5.1968768 & -3.9310157\\
-7.175313 & -6.8201546 & -0.62906033 & 2.0843824 & -5.2005432 & -2.7183036 & -1.2745542 & 1.1503635\\
20.862433 & 19.184505 & -1.5202991 & -3.1956922 & -2.480192 & 0.010624945 & 3.7007527 & 2.4594603\\
7.0845249 & 4.7165266 & 0.43219732 & -4.2952773 & -7.5974342 & -2.0037384 & 3.1573676 & 8.7907618\\
-64.328546 & -63.523828 & 0.78741103 & -15.494941 & -12.659212 & -0.8366425 & 19.738609 & 13.672629\\
6.4134444 & 1.8267049 & -2.8176156 & 4.91351 & 20.662964 & -0.37330178 & -10.477384 & 9.327477\\
65.93701 & 65.223835 & -0.90004386 & -3.6397098 & -2.9529805 & -0.18886984 & 4.5887706 & 3.207476\\
40.204736 & 34.813309 & -5.4050049 & 0.29920049 & 0.21557488 & -0.0010200251 & -0.32442703 & -0.17976199\\
20.600209 & 20.081596 & 0.41455639 & -7.0566527 & -7.2266401 & -0.20823667 & 10.978269 & 8.5102877\\
-20.482695 & -18.717276 & 1.6166552 & -3.3415423 & -2.57432 & 0.015813208 & 3.8363283 & 2.520433\\
123.3774 & 122.07677 & -1.52729 & -9.3685206 & -7.4665641 & -0.63593297 & 11.553057 & 8.1066735\\
-1.9329479 & -0.95373 & -0.70341813 & -2.3953775 & 16.121793 & 4.464689 & -14.534989 & -12.312116\\
-120.01857 & -118.69376 & 1.4582815 & -14.384763 & -11.541028 & -0.93543663 & 17.901586 & 12.503635\\
65.216127 & 63.648383 & -0.75097661 & -52.091821 & -42.607982 & -2.8920707 & 66.532393 & 45.908741\\
195.54734 & 191.51908 & -2.1452241 & 33.022971 & 27.52403 & 1.5653954 & -43.14771 & -29.453425\\
7.1837566 & 6.8299258 & 0.63479725 & 2.1302686 & -5.0604977 & -2.6742399 & -1.2721418 & 1.0067911
\end{tabular}
\caption{Network biases and weights given to 8 significant figures. \begin{scriptsize}
$B2=
\begin{bmatrix}
-5.6955277\\
-19.715168\\
 25.929231\\
 40.363108\\
-3.5442271\\
\end{bmatrix}$
\end{scriptsize}}\label{table_net_coefficients}
\end{table*}

%\bibliography{references}% Produces the bibliography via BibTeX.

\providecommand{\noopsort}[1]{}\providecommand{\singleletter}[1]{#1}%

\end{document}